\documentclass[reprint,amsmath,amssymb,aps,prb,showkeys]{revtex4-1}

\usepackage{graphicx}
\usepackage{color}
\usepackage{dcolumn}
\usepackage{enumitem}
\usepackage{epstopdf}

\usepackage[hyperfootnotes=false]{hyperref}
\begin{document}


\title{Visualization of Magnetic Flux Structure in Phosphorus-Doped EuFe$_2$As$_2$ Single Crystals}

\author{I.\,S. Veshchunov$^{a,b}$}
\author{L.\,Ya. Vinnikov$^{c}$}
\author{V.\,S. Stolyarov$^{a,c}$}
\author{N. Zhou$^{d}$}
\author{Z.\,X. Shi$^{d}$}
\author{X.\,F. Xu$^{e}$}
\author{S.\,Yu. Grebenchuk,$^{a}$}
\author{D.\,S. Baranov$^{a,c,f}$}
\author{I.\,A. Golovchanskiy$^{a,g}$}
\author{S. Pyon$^{b}$}
\author{Yue Sun$^{b,h}$}
\author{Wenhe Jiao$^{i}$}
\author{Guanghan Cao$^{i}$}
\author{T. Tamegai$^{b}$}
\author{A.\,A. Golubov$^{a,j}$}
\affiliation{$^{a}$Moscow Institute of Physics and Technology (State University), Dolgoprudnyi, Moscow region, 141700 Russia}%
\affiliation{$^{b}$Department of Applied Physics, The University of Tokyo, 7-3-1 Hongo, Bunkyo-ku, Tokyo 113-8656, Japan}%
\affiliation{$^{c}$Institute of Solid State Physics, Russian Academy of Sciences, Chernogolovka, Moscow region, 142432 Russia}%
\affiliation{$^{d}$Department of Physics and Key Laboratory of MEMS of the Ministry of Education, Southeast University,
211189 Nanjing, China}%
\affiliation{$^{e}$Department of Physics, Changshu Institute of Technology, Changshu 215500, People's Republic of China}%
\affiliation{$^{f}$Laboratoire de physique et d’etude des materiaux, LPEM-UMR8213/CNRS-ESPCI ParisTech-UPMC,
75005 Paris, France}%
\affiliation{$^{g}$National University of Science and Technology MISIS, Moscow, 119049 Russia}%
\affiliation{$^{h}$Institute for Solid State Physics, The University of Tokyo, 277-8581 Kashiwa, Japan}%
\affiliation{$^{i}$Department of Physics, Zhejiang University, 310027 Hangzhou, China}%
\affiliation{$^{j}$Faculty of Science and Technology and MESA+ Institute of Nanotechnology, University of Twente, 7500 AE Enschede, The Netherlands}%

\begin{abstract}
Magnetic flux structure on the surface of EuFe$_2$(As$\rm_{1-x}$P$\rm_x$)$_2$ single crystals with nearly optimal
phosphorus doping levels $x=0.20$, and $x=0.21$ is studied by low-temperature magnetic force microscopy and decoration with ferromagnetic nanoparticles. The studies are performed in a broad temperature range. It is shown that the single crystal with $x=0.21$ in the temperature range between the critical temperatures $T_{\rm SC}=22$~K and $T_{\rm C}=17.7$~K of the superconducting and ferromagnetic phase
transitions, respectively has the vortex structure of a frozen magnetic flux, typical for type-II superconductors. The magnetic
domain structure is observed in the superconducting state below $T_{\rm C}$. The nature of this structure is discussed.
\end{abstract}

\keywords{ferromagnetic superconductor; spontaneous vortex phase; magnetic force microscopy}


\maketitle

\section{Introduction}

The coexistence of superconductivity and magnetic ordering has been a subject of a strong interest.\cite{Huxley} Currently, the electric transport and magnetic properties are well studied for a number of single-crystalline compounds of the so-called magnetic superconductors: borocarbides,\cite{Gupta,Canfield,Canfield1,Kawano,Chia,Gammel} uranium compounds,\cite{Huy} high-temperature cuprate superconductors,\cite{Lavrov_PRB_79_214523} and iron-based superconductors.\cite{Kamihara} 

An important issue of the coexistence of superconductivity and magnetism from both theoretical \cite{Ng,Buzdin1,Faure,Sonin} and experimental perspectives \cite{Iavarone,Tamegai,Xing,Yang} relates to the microstructure of the magnetic flux, as well as to its dynamics upon variation of the temperature and external magnetic field. Until recently, low temperatures of superconducting and magnetic phase transitions of known single crystals, as well as the requirement of a high spatial resolution, have limited experimental capabilities for visualization of the magnetic flux structure employing e.g. magnetic force microscopy (MFM),\cite{Bluhm} scanning Hall probe imaging,\cite{Bluhm1} and decoration with magnetic nanoparticles. \cite{Vinnikov_Springer_23_89,March,decor} 

Recently, new iron-based compounds AFe$_2$(As$_{1-x}$P$_x$)$_2$ (where А = Ba, Sr, Ca, Eu) have been synthesized. Superconductivity in these compounds can be induced by doping with phosphorus.\cite{Ren} Superconductivity in EuFe$_2$(As$_{1-x}$P$_x$)$_2$ single crystals occurs in a rather narrow doping range $x = 0.14-0.25$ (or in the phosphorus content range $7.0-12.5$ at\%) with the maximum superconducting transition temperature $T_{\rm SC}^{\rm max}=27$~K.\cite{Jeevan,Adachi} The magnetic transition in the Eu$^{2+}$ subsystem is observed at temperatures $T_{\rm C}\sim17-20$~K and depends moderately on the phosphorus content (doping level) in the specified range of contents.\cite{Jeevan,Nandi1,Zapf,Zapf1,Pogrebna,Goltz,Kadowaki}

Previously, the magnetic flux structure was visualized with the MFM on artificial thin-film superconductor/ferromagnet (Nb/FeNi) hybrid structures,\cite{Iavarone} where the domain structure and Abrikosov vortices frozen in the superconductor were observed simultaneously. However, in Ref.~\onlinecite{Iavarone} the Curie temperature $T_{\rm C}$ of ferromagnetic layers was much higher than the critical temperature of the superconducting transition $T_{\rm SC}$ in niobium films. Also, vortex structures were observed in spatially homogeneous ErNi$_2$B$_2$C bulk superconducting single crystals ($T_{\rm SC}=10.5$~K) in Ref.~\onlinecite{Veschunov} using the decoration method, and interpreted as an evidence of presence of domain boundaries in a weakly ferromagnetic phase with $T_{\rm C}=2.3$~K.

In this work, the structure of the magnetic flux in EuFe$_2$(As$_{1-x}$P$_x$)$_2$ single crystals with $x=0.20$ and $x=0.21$ is studied with MFM and Bitter decoration technique in a broad temperature range. Stripe and maze domain structures typical for ferromagnets with perpendicular magnetic anisotropy, are observed in the superconducting state below $T_{\rm C}$. In contrast to artificial hybrid systems, in EuFe$_2$(As$_{1-x}$P$_x$)$_2$ an interface is absent and superconductivity and ferromagnetism coexist on the atomic scale.
\section{Experimental details}

EuFe$_2$(As$_{1-x}$P$_x$)$_2$ single crystals were synthesized using the self-flux method.\cite {Zhou} The actual composition of synthesized single crystals was determined by energy dispersive X-ray (EDX) microanalysis employing Carl Zeiss Supra 50 VP SEM microscope. For MFM and decoration studies, single crystals of EuFe$_2$(As$_{0.8}$P$_{0.2}$)$_2$ and EuFe$_2$(As$_{0.79}$P$_{0.21}$)$_2$ of $1\times1\times0.012$ mm$^3$ size with an atomically smooth surface were obtained by mechanical cleavage. Temperature and field dependences of the magnetization were measured on Quantum Design MPMS-XL5 SQUID magnetometer at fields up to 5 T. The surface structure and the distribution of magnetic flux were studied using AttoCube AttoDry 1000 atomic force microscope (AFM) with a closed-cycle cryogenic system, and a base temperature of 4 K. For AFM and MFM studies silicon cantilevers were used coated by magnetic CoCr layer (MESP, Bruker) with the following characteristics at 4.2 K: the resonance frequency of the cantilever 87 kHz, the stiffness constant 2.8 N/m, and the coercive field $\approx1400$ Oe. AFM/MFM imaging was performed in an atmosphere of exchange gas (helium) at pressure $P\sim0.5$ mbar in the temperature range from 4 to 30 K, controlled with exceptional precision of 1 mK. Prior to MFM imaging, probes were magnetized at $H = 2$ kOe above the superconducting transition temperature $T_{\rm SC}=22$~K of EuFe$_2$(As$_{0.79}$P$_{0.21}$)$_2$ sample. The topography of the surface was studied in the tapping mode and magnetic flux structure was imaged in the MFM lift mode at 110 nm above the sample surface with the feedback switched off and fast scanning direction along the $Y$ axis. The MFM contrast was provided by the phase shift in the cantilever oscillation. The decoration of the surface of EuFe$_2$(As$_{0.8}$P$_{0.2}$)$_2$ single crystal was performed with magnetic iron particles ($\sim10$ nm) in the field cooling (FC) regime at liquid helium temperatures.\cite{Vinnikov_Springer_23_89} 

\section{Results}
%

\begin{figure}[ht!]
\includegraphics[width=0.70\columnwidth]{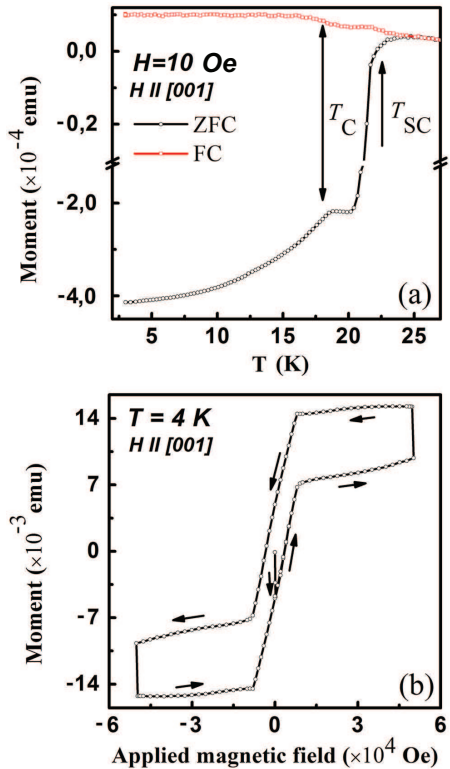}
\caption{(Color online) Temperature dependence of the
magnetization and magnetization curve for  EuFe$_2$(As$_{0.79}$P$_{0.21}$)$_2$ single crystal. (a) Temperature dependence of the magnetization measured in the FC regime with the field parallel to the $c$-axis of the crystal and in the zero-field cooling (ZFC) regime. Transitions to the superconducting and ferromagnetic states are observed at $T_{\rm SC}=22$~K and $T_{\rm C}=(18\pm0.3)$~K, respectively (marked by arrows). (b) The dependence of the magnetization on the applied magnetic field at $T=4$ K.} 
\label{MTH}
\end{figure}

Fig.~\ref{MTH} shows typical magnetic properties of EuFe$_2$(As$_{0.79}$P$_{0.21}$)$_2$ single crystal. Fig.~\ref{MTH}(a) demonstrates temperature dependences of the magnetization measured in the FC and zero-field cooling (ZFC) regimes. The superconducting transition temperature $T_{\rm SC}=22$~K is indicated by the right arrow. Step features on the ZFC and FC temperature dependences
of the magnetization are attributed to a ferromagnetic phase transition. It is noteworthy that a transition to the superconducting state is also accompanied by the appearance of residual magnetization upon cooling in
an external field of 10 Oe. Fig.~\ref{MTH}(b) shows the dependence
of the magnetization on the applied magnetic field parallel to the $c$-axis of the crystal. For the sample with $x=0.20$ the temperature
dependence of the magnetization and magnetization curve at 4 K are similar but with a higher superconducting transition temperature and a wider hysteresis loop.

\begin{figure*}[ht!]
\includegraphics[width=0.63\columnwidth]{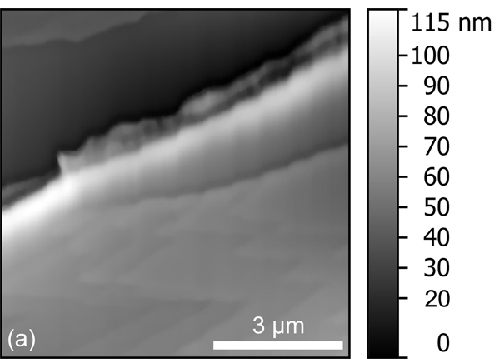}
\includegraphics[width=0.65\columnwidth]{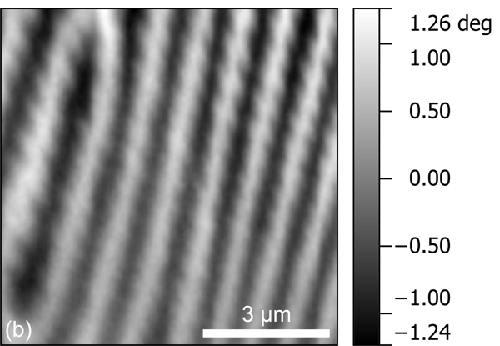}
\includegraphics[width=0.66\columnwidth]{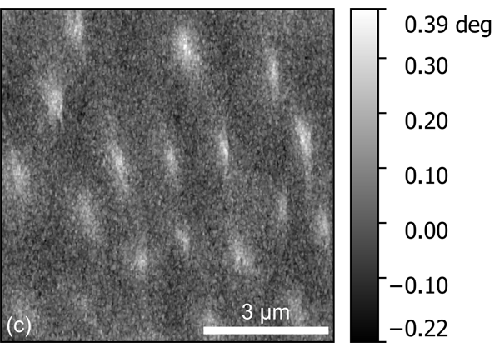}
\caption{AFM topographic image and MFM images of magnetic flux structure on the (001) surface of EuFe$_2$(As$_{0.79}$P$_{0.21}$)$_2$ single crystal. (a) AFM topography of the surface area in fully magnetized state of the Eu$^{2+}$ ferromagnetic subsystem in magnetic field of $H=-0.9$ T parallel to the $c$-axis. (b) Magnetic domain structure after zero-field cooling (ZFC) to the minimum temperature $T_{\rm min}=4.16$~K with subsequent heating up to $T=17.27$~K. (c) Vortex structure imaged after FC at $T=18.15$~K with the residual magnetic flux density $\Phi_0/a^2\sim6$~G.} 
\label{AFM/MFM}
\end{figure*}

Fig.~\ref{AFM/MFM} shows the results of the AFM/MFM studies. Fig.~\ref{AFM/MFM}(a) demonstrates the AFM topography of the $ 8\times8\,\rm\mu m^2$ surface area of EuFe$_2$(As$_{0.79}$P$_{0.21}$)$_2$ single crystal with the step of $\sim100$ nm height. Fig.~\ref{AFM/MFM}(b) shows the distribution of the
magnetic flux over the surface shown in Fig.~\ref{AFM/MFM}(a) at $T=17.27$~K. This structure is typical for the entire temperature range below the Curie temperature and disappears after heating
above $T_{\rm C}$. Thus, the observed sign-alternating contrast can be attributed to the magnetic domain structure. Importantly, the domain structure is observed not only at zero external magnetic field, but also upon cooling in weak fields $H<100$~Oe. Fig.~\ref{AFM/MFM}(c) shows the distribution of the magnetic flux in the superconducting state in a
narrow temperature range above $T_{\rm C}$. The observed
contrast (light spots) corresponds to Abrikosov vortices with the magnetic flux density $\Phi_0/a^2\sim6$~G, where $\Phi_0$ is the magnetic flux quantum and $a$ is the average distance between vortices.

\begin{figure*}[ht!]
\includegraphics[width=0.7\columnwidth]{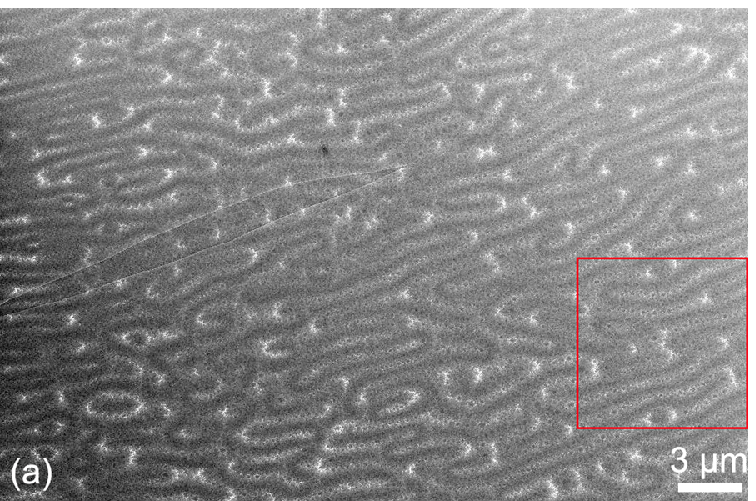}
\includegraphics[width=0.46\columnwidth]{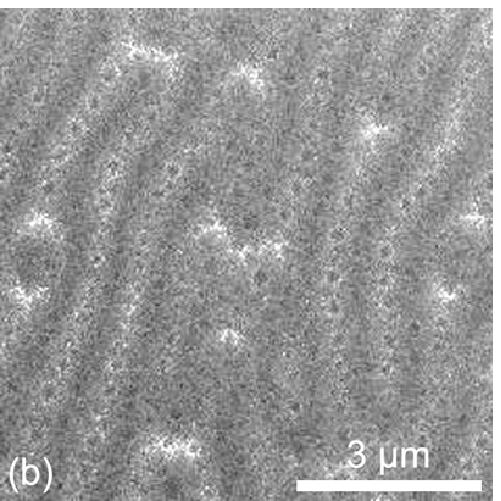}
\caption{(Color online) (a) Magnetic flux structure on the (001) surface of EuFe$_2$(As$_{0.80}$P$_{0.20}$)$_2$ single crystal after FC in a magnetic field of 10 Oe revealed by decoration at $T_{\rm d}\sim8$~K. (b) Enlarged image of the area indicated by the red box in (a) of similar size and the orientation of the domain structure shown in Fig. 2b.} 
\label{Bitter}
\end{figure*}

Fig.~\ref{Bitter} shows the typical magnetic flux structure observed by the decoration method on the (001) surface of EuFe$_2$(As$_{0.80}$P$_{0.20}$)$_2$ single crystal with the superconducting transition temperature $T_{\rm SC}=24$~K. With MFM only a small $\sim8\times8\,\rm\mu m^2$ surface area of the sample was studied, whereas the decoration method reveals the magnetic structure on the almost entire surface. According to the principle of the image contrast formation in the decoration method,\cite{Veshchunov_JETP_88_758,Sakurai} the region of higher density of magnetic particles (light) is treated as a domain with the magnetization along the applied field direction, whereas the region with lower density or without magnetic particles (dark) is interpreted as a domain with the opposite sign of the magnetization. As can be seen, the decorated domain structure (Fig.~\ref{Bitter}(b)) agrees with MFM imaged one (Fig.~\ref{AFM/MFM}(b)) at corresponding scales. The period of the domain structure is about 0.9 $\rm\mu m$. At the same time, finer details of the decorated domain structure can be resolved (Fig.~\ref{Bitter}(b)). The magnetization measurements and both MFM imaged and decorated magnetic structure define explicitly EuFe$_2$(As$_{0.79}$P$_{0.21}$)$_2$ and EuFe$_2$(As$_{0.80}$P$_{0.20}$)$_2$ single crystals as superconductors with ferromagnetic ordering and the superconducting transition temperature $T_{\rm SC}$ above the Curie temperature $T_{\rm C}$.

\section{Discussion}
The experimental results can be interpreted as follows.
According to the dependences shown in Fig.~\ref{MTH}(a), the ZFC magnetization is negative below the superconducting transition temperature
$T_{\rm SC}=22$~K. In the temperature range below $T_{\rm C}$, the diamagnetic response is weakened by the ferromagnetic transition in the Eu$^{2+}$ subsystem. The exact determination of the Curie temperature using the observed features on the ZFC and FC temperature dependences of the magnetization is complicated due to competing mechanisms of superconducting and ferromagnetic orderings. In particular, maxima on
the FC and ZFC temperature dependences of the magnetization are observed in Ref.~\onlinecite{Jeevan} at $T\sim17.7$~K, whereas according to measurements of the specific heat, the Curie temperature is $T_{\rm C}=19$~K. In this work, the transition temperature to the ferromagnetic state $T_{\rm C}$ is defined as a temperature at which the domain structure is first observed, i.e., $T_{\rm C}=17.7$~K. The dependence of the magnetization on the applied magnetic field (Fig.~\ref{MTH}(b)) is the superposition
of a typical hysteresis loop of a type-II superconductor (within the Bean model the critical current density $J\rm_c$ is proportional to the width of the hysteresis loop) and the magnetization curve of the Eu$^{2+}$ ferromagnetic subsystem.\cite{Jeevan} 

The magnetic origin of the domain structure contrast (Figs.~\ref{AFM/MFM}(b) and \ref{Bitter}(a)) is confirmed by insensitivity of the MFM probe to small details of the surface topography, e.g. to the 100 nm step. Sign-alternating (phase) contrast on domains indicates perpendicular magnetic anisotropy and corresponds to
the antiparallel direction of the magnetization in neighboring domains. 

Individual vortices could not be resolved with decoration since the expected magnetic flux density within domains is about 0.9 T ($M\rm_s = 714$ cgs units/cm$^3$) at liquid helium temperatures,\cite{Jeevan} whereas the resolution of the decoration method is limited by 0.2 T.\cite{Vin3} The spatial resolution of MFM also cannot identify individual vortices if
the local magnetic flux density in domains is much higher than 10 mT.\cite{Volodin,Auslaender} At the same time, the fine structure of domains, which is shown in Fig.~\ref{Bitter}(b), can be explained within the framework of domain branching in ferromagnets.\cite{Hubert} An alternative origin of the fine domain structure is the so-called intermediate-mixed state,\cite{Golubok_JETP_35_642} which
appears if the thickness of the superconducting crystal is much larger than the width of domains and is characterized by a mixture of flux-free domains (Meissner phase) and domains with Abrikosov vortices. In
contrast to the structure of the intermediate-mixed state, the fields of vortices in neighboring branching domains should be oppositely oriented. Such a possibility was theoretically considered in Ref.~\onlinecite{Faure}. According to this model, different types of domain configurations can be formed in a ferromagnetic superconductor depending on the parameters
(magnetic and superconducting): the saturation magnetization ($M\rm_s$), London penetration depth ($\lambda$), the lower critical field ($H_{\rm c1}$), and the domain wall width $w$. Precise measurements of these parameters and studies of the fine structure of domains will provide further clarification of the mechanisms of the coexistence and mutual influence of superconductivity and ferromagnetism in EuFe$_2$(As$_{1-x}$P$_x$)$_2$ single crystals.
\section{Concluding remarks}

The main result of this work is the observation of
the magnetic domain structure in EuFe$_2$(As$_{1-x}$P$_x$)$_2$ superconducting single crystals with $x = 0.20$, and $x = 0.21$. This domain structure disappears in EuFe$_2$(As$_{0.80}$P$_{0.20}$)$_2$ single crystal after heating above the Curie temperature $T_{\rm C}=17.7$~K. Thus, the magnetic domain structure has been observed for the
first time in spatially homogeneous single crystals with the superconducting transition temperature $T_{\rm SC}$ exceeding the Curie temperature $T_{\rm C}$, 
$T_{\rm SC}>T_{\rm C}$,
which unambiguously indicates the coexistence of ferromagnetism and superconductivity in this material.
The observations of the magnetic flux structure using low-temperature MFM and decoration methods in real space (in contrast to X-ray and neutron diffraction studies) provide important information on the topology, real sizes, and shape of domains. At the same time, only further
combined studies employing, e.g. diffraction methods and scanning probe microscopy, in particular high-resolution scanning tunneling
microscopy, as well as decoration with ferromagnetic nanoparticles in a broad range of temperatures and magnetic fields can clarify the mechanism of the coexistence of superconductivity and ferromagnetism in these ferromagnetic superconductors.

\section{Acknowledgments}
I.S.V, L.Ya.V., and V.S.S. are grateful to V.V. Ryazanov, L.S. Uspenskaya, S.I. Bozhko, and A.I. Buzdin for stimulating discussions. We are grateful to V.V. Dremov, E.Yu. Postnova, A.G. Shishkin, and
L.G. Isaeva for assistance. N.Z., Z.X.S., X.F.X., W.J., and G.C. acknowledge the support by the National Science Foundation of China (No. 11474252, 11611140101
and U1432135). The work of V.S.S., D.S.B., and I.A.G. was supported by the Russian Foundation for Basic Research (project No. 16-32-60133 mol-a-dk and 16-32-00309 mol-a). The MFM studies were supported by the Ministry of Education and Science of the Russian Federation (project No. 14.Y26.31.0007).


\end{document}